\documentclass[twocolumn]{aastex631}

\usepackage{amsmath}
\usepackage{booktabs}

\usepackage{array}
\usepackage{lipsum}

\begin{document}

\title[Probing Baryon Spread with FRB]{Constraining Baryonic Feedback Effects on the Matter Power Spectrum with Fast Radio Bursts}

\correspondingauthor{Isabel Medlock}
\email{isabel.medlock@yale.edu}

\author{Isabel Medlock}
\affiliation{Department of Astronomy, Yale University, New Haven, CT 06520, USA}

\author{Daisuke Nagai}
\affiliation{Department of Physics, Yale University, New Haven, CT 06520, USA}
\affiliation{Department of Astronomy, Yale University, New Haven, CT 06520, USA}

\author{Daniel Anglés-Alcázar}
\affiliation{Department of Physics, University of Connecticut, 196 Auditorium Road, U-3046, Storrs, CT, 06269, USA}

\author{Matthew Gebhardt}
\affiliation{Department of Physics, University of Connecticut, 196 Auditorium Road, U-3046, Storrs, CT, 06269, USA}

\begin{abstract}

In the age of large-scale galaxy and lensing surveys, such as DESI, Euclid, Roman and Rubin, we stand poised to usher in a transformative new phase of data-driven cosmology. To fully harness the capabilities of these surveys, it is critical to constrain the poorly understood influence of baryon feedback physics on the matter power spectrum. We investigate the use of a powerful and novel cosmological probe - fast radio bursts (FRBs) - to capture baryonic effects on the matter power spectrum, leveraging simulations from the CAMELS projects, including IllustrisTNG, SIMBA, and Astrid. We find that FRB statistics exhibit a strong correlation, independent of the subgrid model and cosmology, with quantities known to encapsulate baryonic impacts on the matter power spectrum, such as baryon spread and the halo baryon fraction. We propose an innovative method utilizing FRB observations to quantify the effects of feedback physics and enhance weak lensing measurements of $S_{8}$. We outline the necessary steps to prepare for the imminent detection of large FRB populations in the coming years, focusing on understanding the redshift evolution of FRB observables and mitigating the effects of cosmic variance.

\end{abstract} 

\section{Introduction} \label{sec:intro}

In the era of precision cosmology, current and upcoming galaxy and lensing surveys - such as DESI \citep{DESI_2016}, Euclid \citep{Euclid_2022}, Roman \citep{WFIRST_2015}, and Rubin \citep{Ivezic_2019A} - are poised to transform our understanding of the Universe. A central scientific objective for modern cosmology in the forthcoming decade is to harness these surveys to significantly enhance the statistical power of cosmological measurements. This will enable more precise constraints on the nature of cosmic structure and its evolution over time.

One parameter of particular interest and contention is $S_{8}$, which describes the amplitude of matter fluctuations on the scale of 8 Mpc/h, or in other words the 'clumpiness' of the Universe. Intriguingly, a 2-3 sigma tension exists between current measurements of $S_{8}$, between late Universe probes such as galaxy clustering, weak lensing, and the thermal Sunyaev-Zel'dovich (SZ) effect and early Universe measurements from the cosmic microwave background \citep[e.g.,][]{PlanckParams_2020, KidsParams_2021, DES_2022a, DES_2022b, DES_2023}. Although this tension is still only a few sigma, it remains tantalizing and hints at significant implications for the potential discovery of new physics.

A plausible explanation for this tension involves baryonic effects, which when neglected, can result in biases greater than statistical errors in cosmological parameter measurements \citep[e.g.,][]{Semboloni_2011, Zentner_2013}. More broadly, the impact of baryonic physics on the matter power spectrum is considered one of the primary theoretical uncertainties that must be understood to a precision of a few percent or better to maximize the statistical power of the upcoming survey data. For instance, current and future surveys leveraging weak lensing to derive $\sigma_{8}$, and $S_{8}$ require predictions of the matter power spectrum with accuracy better than 1\% at $k < $10 $h$/Mpc \citep{Huterer_2005, Laureijs_2009, Ivezic_2019A}.
 
The matter power spectrum, $P(k)$, is defined as the squared amplitude of matter overdensities, as a function of the Fourier scale ($k =2\pi/\lambda$ where $\lambda$ is the physical scale). Baryonic feedback effects, such as radiative cooling, star formation, and feedback from galactic winds and active galactic nuclei (AGN), redistribute gas and thereby modify the matter power spectrum compared to a dark matter-only universe \citep[e.g.,][]{vanDaalen2011, Mummery_2017, springel_illustris_clustering_2018, Chisari_2018, Chisari_2019, vanDaalen_2020, Delgado_2023, Salcido_2023, Hernandez-Aguayo_2023, Gebhardt_2024, Schaller_2024}. On larger scales, baryons suppress power as feedback mechanisms push matter out of halos. Conversely, on small scales, power is enhanced as cooling and star formation increase the clustering of matter. This redistribution effect is significant across certain scales. For example, \cite {vanDaalen2011} find that the redistribution of matter due to stellar and AGN feedback reduces the power relative to a dark-matter-only Universe by 1\% at scales of $k= 0.3$~$h$/Mpc but reaches 28\% at scales of $k = 10$~$h$/Mpc. To a lesser extent, this redistribution of baryons leads to a change in the clustering of dark matter as well, an effect known dark matter back-reaction. 

Despite their substantial influence, we lack a comprehensive understanding of these baryonic feedback processes. These processes are highly complex and non-linear, rendering them unsuitable for analytical modeling. Additionally, they span a wide range of scales that exceed the resolution of state-of-the-art simulations. Consequently, simulations must rely on subgrid models, which, despite being calibrated on common observables, such as the stellar-to-halo mass relation, can differ greatly from one another \cite[e.g.,][for a review]{Vogelsberger_2020}. Different implementations of these feedback models can lead to significantly divergent predictions, particularly for the matter power spectrum \citep[e.g.,][]{Delgado_2023, Gebhardt_2024}, and the distribution of gas around galaxies \citep[e.g.,][]{Chisari_2018, Chisari_2019, vanDaalen_2020}. 

The current approach to modeling these baryonic effects involve marginalizing over the relevant parameters. Several methods have been developed to achieve this, including compiling suites of hydrodynamical simulations varying feedback \citep[e.g.,][]{OWLS_2014, BAHAMAS_2017, camels_presentation, Flamingo_2023}, parameterizing halo profiles \citep[e.g.,][]{Mead_2015, Mead_2016} and applying baryonification techniques to N-body simulations \citep[e.g.,][]{Schneider_2015, Schneider_2019}. A recently proposed approach suggests that observed halo baryon fractions and retained halo mass are promising tracers of the extent of baryonic suppression on the matter power spectrum \citep{vanLoon_2024}. Another recent study identified that the baryon spread metric - which measures how far baryons spread from their host halos on average - is also promising in capturing the effects of feedback physics on the matter power spectrum \citep{Gebhardt_2024}. However, these quantities are either impossible or very challenging to measure observationally, underscoring the need for a simple and straightforward observable alternative. 

We demonstrate that fast radio bursts (FRBs) are a promising probe for characterizing and mitigating the effects of baryons on the matter power spectrum. FRBs, intriguing phenomena in their own right, are excellent tracers of baryons \citep{Fujita_2017, Ravi_2019, Battaglia_2019}. In previous work, we utilized the CAMELS project \citep{camels_data_release, camels_presentation, camels_data_release2} to show that FRB statistics have the potential to distinguish between different feedback models. Recently, \cite{Guo_2025} built upon this work with a larger set of simulations from the CAMELS project, corroborating the future constraining power of FRBs. With rapidly advancing observational capabilities - such as CHIME outriggers \citep{Leung_2021}, CHORD \citep{Vanderline_2019}, DSA-2000 \citep{Hallinan_2019}, and BURSST \citep{BURSTT_2022} - FRBs represent a key avenue to pursue in order to maximize the power of galaxy and lensing surveys for cosmological constraints. 

The structure of this paper is as follows: In Section~\ref{sec:frbs} we review relevant background information about FRBs.  Section~\ref{sec:met} provides a summary of the simulations used and the methods to calculate the parameters. Our main results are presented in Section~\ref{sec:res}.
In Section~\ref{sec:disc}, we discuss the scientific potential, next steps, and future directions. Finally, in Section~\ref{sec:conc}, we summarize our conclusions, key takeaways, and provide an outlook for the coming years.

\section{Fast Radio Bursts} \label{sec:frbs}

FRBs are millisecond long, extremely energetic pulses of radio waves of mysterious origins propagating from cosmological distances \citep{Lorimer_2007, Thornton_2013}. First discovered in archival pulsar data in 2007 \citep{Lorimer_2007}, FRBs are now detected in their hundreds and thousands \citep{CHIME_2021}. 

FRBs offer several advantages over other traditional probes. FRBs provide a direct measurement of the column density of free electrons along a line of sight without dependency on temperature like the thermal SZ effect. FRBs are also highly energetic and thus detectable across a wide range of cosmological distances and are frequent events, with thousands of events per day detectable from Earth. Using these features to their advantage, FRBs have lent insight to a vast array of problems across the fields of cosmology and galaxy physics. Recently, FRBs were used to locate missing baryons \citep{McQuinn_2014, Macquart_2020}. Additionally, FRBs have been shown to be promising probes of the structure of galaxy halos and the circumgalactic medium (CGM), the intergalactic medium (IGM), and large-scale structure \citep{Connor_2022, Wu_2023, Walker_2023} as well as the baryonic feedback processes that shape these structures \citep{Baptista_2023, Medlock_2024}. FRBs also have the power to constrain parameters such as $\Omega_{\rm b}$, $\Omega_{\rm m}$ and $H_{0}$ \citep{Macquart_2020, Nicola_2022, James_2022b}. 

The key observable associated with FRBs is the dispersion measure (DM). The DM is related to the time delay waves of different frequency experience as they interact with matter along their paths and thus traces the distribution of ionized baryons along the path of the FRB. DM is defined as the integrated electron density along the line of sight (LOS) from the source to the observer,  
\begin{equation} \label{eq:dmdef} \centering
\rm DM = \int_{0}^{d} \frac{n_{e}(l)}{1+z}dl,
\end{equation}
\noindent where \emph{d} is the proper distance, n\textsubscript{e} is the free electron number density, \emph{z} is the redshift, and \emph{l} is the proper path length. The Milky Way, FRB host galaxy, and the IGM and CGM of intervening halos all contribute to the dispersion, so that the observed DM is the sum of all of these contributions.

\begin{equation} \label{eq:dmbreakdown} \centering
\rm DM_{obs} = DM_{MW} + DM_{IGM} + DM_{CGM} + \frac{DM_{Host}}{1+z}.
\end{equation}

The relationship between redshift and dispersion measure from extragalactic contributions ($\rm DM_{cosmic} = DM_{IGM} + DM_{CGM}$) has been well defined, derived from theory, and confirmed through observations and simulations \citep{Macquart_2020, Zhang_2021, James_2022, Yang_2022}. Assuming a flat cosmology, the expected DM from extragalactic sources (IGM, intervening halos), dubbed the Macquart relation, is given by:

\begin{equation} \centering
\rm \langle DM_{cosmic} \rangle = \int^{z}_{0} \frac{c f_d \rho_b(z)m_p^{-1} (1 - Y_{He}/2)dz}{H_0(1+z)^2\sqrt{\Omega_m (1+z)^3 + \Omega_{\Lambda}}},
\label{eq:mac_rel}
\end{equation}

\noindent where $f_{\rm d}$ is the fraction of cosmic baryons in diffuse ionized gas, $\rm \rho_b(z) = \Omega_b\rho_{c,0}(1+z)^3$, $m_p$ is the proton mass, $Y_{\rm He}$ is the mass fraction of helium assumed doubly ionized, and $\Omega_{\rm m}$ and $\Omega_{\rm \Lambda}$ are the total matter and dark energy densities at the present day in units of the critical density. Thus, the exact calibration of this relation depends on the values of these cosmological parameters. At a given $z$, the probability of deviations from $\rm \langle DM_{\rm cosmic} \rangle$ is given by 

\begin{equation} \label{eq:p_dm}
    p_{\rm cosmic}(\Delta) = A \rm \Delta^{-\beta}exp\left[-\frac{(\Delta^{-\alpha} - C_0)^2}{2\alpha^2\sigma_{DM}^2}\right], 
    %\: \: \Delta > 0, 
\end{equation}

\noindent where $\rm \Delta = DM_{\rm cosmic}/\langle DM_{\rm cosmic} \rangle >0$, $\rm \alpha = \beta = 3$, $C_{0}$ is tuned so that the expectation value is unity, and $\rm \sigma_{DM}$ describes the spread of the distribution. The standard deviation of this distribution $\rm \sigma_{DM}$ can be described as

\begin{equation}
    {\sigma_{\rm DM}(\Delta)}  = Fz^{-1/2},
    \label{eq:fparam}
\end{equation}

\noindent where $\sigma_{\rm DM}(\Delta)$ is the fractional standard deviation of the dispersion measure, $z$ is the redshift (the $z^{-1/2}$ scaling is due to the Poisson nature of the intersecting halos), and $F$ is the F-parameter \citep{McQuinn_2014, Macquart_2020}. Thus, $F$ captures the degree of deviation from $\langle \rm DM_{\rm cosmic} \rangle$, and the strength of the baryonic feedback. In cases of stronger feedback, baryons are pushed further from the halos. This leads to a more uniform distribution of baryons, less deviation from $\langle \rm DM_{\rm cosmic} \rangle$ and a lower value of $F$, approaching zero. In the case of weaker feedback, baryons remain close to their halos, leading to a greater deviation from $\langle \rm DM_{\rm cosmic} \rangle$, and a higher value for $F$. Since $F$ is sensitive to the overall distribution of baryons, it also captures the effect of the large-scale structure that is not sensitive to feedback but encodes cosmological information.

\section{Methods} \label{sec:met}

\subsection{The CAMELS Project}

The Cosmology and Astrophysics with MachinE Learning Simulations\footnote{\href{https://www.camel-simulations.org/}{https://www.camel-simulations.org/}} (CAMELS) project is the largest collection of cosmological hydrodynamical simulations, currently consisting of over 14,000 simulations, including nearly 8000 (magneto)hydrodynamic simulations and 6,000 N-body simulations. With this massive collection of simulations, the CAMELS project aims to enhance our understanding of baryonic feedback physics effects through the use of machine learning to marginalize over uncertainties in simulations \citep{camels_presentation, camels_data_release}. Currently, CAMELS comprises ten different simulation suites using a variety of codes and 9 subgrid models: IllustrisTNG, SIMBA, Astrid, Magneticum, SWIFT-EAGLE, Ramses, Enzo, Obsidian, CROCODILE, and N-Body. In addition to varying the code and subgrid model used, hydrodynamic simulations vary feedback and cosmology parameters, resulting in thousands of simulations with different models of baryonic feedback physics. All simulations follow the evolution of $\rm 256^3$ dark matter particles (with mass $6.49 \times 10^{7} (\Omega_{\mathrm{m}} - \Omega_{\mathrm{b}})/0.251 h^{-1} M_\odot$) and, for hydrodynamical simulations, $\rm 256^3$ fluid elements (with initial mass $1.27 \times 10^{7} h^{-1} M_\odot$), from $\rm z = 127$ to $\rm z = 0$ in a periodic box with sides of comoving length $\rm L = 25$ $h^{-1}$~Mpc. Most simulations are run with the following cosmological parameters: $\rm \Omega_{b} = 0.049$, $h = 0.6711$, $\rm n_s = 0.9624$, $\Sigma m_{\rm s} = $0.0~eV, and $w = -1$, with the exception of those that vary these parameters.

For our analysis, we used the CAMELS-SIMBA, CAMELS-IllustrisTNG, and CAMELS-Astrid suites. The CAMELS-SIMBA suite contains 1,171 hydrodynamic simulations run with the GIZMO code \citep{Hopkins_2015} and the same subgrid physics as the original SIMBA simulations \citep{dave_simba_2019}. The CAMELS-IllustrisTNG suite is made up of 3,219 hydrodynamic simulations run with the AREPO code \citep{Springel_2010, Weinberger_2020} and the same subgrid physics as the original IllustrisTNG simulations \citep{Weinberger_2017, Pillepich_2018}. CAMELS-Astrid includes 2,080 hydrodynamic simulations performed with the MP-Gadget code, a highly scalable version of Gadget-3 \citep{Springel_2005}, and a subgrid model based on the original Astrid code described in \citet{Bird_2022, Ni_2022}.
See \cite{Medlock_2024b} for a more in-depth discussion of the differences in implementation between the IllustrisTNG and SIMBA subgrid models. For further information on the CAMELS project, we refer the reader to \citet{camels_data_release, camels_presentation, camels_data_release2}.

\subsection{Calculating the F-Parameter}

The dispersion measure distributions and the F-parameter values of the CAMELS SIMBA, IllustrisTNG, and Astrid simulation suites were calculated and described in detail in our previous paper \citep{Medlock_2024}. 

DM sightline calculations start with DM column density maps over single simulation boxes at the available redshifts. Using the electron density field, yT \citep{turk11} creates a $4000\times4000$ pixel grid. For the $L = 25$~comoving $h^{-1}$Mpc boxes this results in pixels that are 9.3~comoving kpc across, and represents the electron column density integrated along the depth of the simulation box. Following Equation~\ref{eq:dmdef} we divide the electron column density by $(1+z)$ and convert units from the column density (cm$^{-2}$) to DM (pc/cm$^{3}$).

To construct a single sightline out to the cosmological distance, we stitch simulation boxes together to reach the desired distance. We keep a running sum of the distance our sightline has traveled that we can use to determine which redshift snapshot is most appropriate to use and to determine when we have reached our desired distance. To avoid repeated structure, we choose a random pixel from the map of each box we add. For each simulation, we repeat this sightline construction to form a distribution of 10,000 sightlines integrated to $z=0.25, \hspace{1pt} 0.5, \hspace{1pt}0.75, \hspace{1pt}\rm{and} \hspace{1pt}1.0$.

For each simulation, we calculate $F$ using Equation~\ref{eq:fparam}. We take $\sigma_{\rm DM}(\Delta)$ as the standard deviation of the DM of all 10,000 sightlines divided by the mean and $z$ as the redshift in which these sightlines were integrated. This results in a single dimensionless number that characterizes the width of the DM distribution of an entire simulation (for a given $z$). The error is then calculated through a bootstrapping analysis.  We calculate $F$ for each simulation in the one-parameter (1P) set of SIMBA, Astrid, and IllustrisTNG. In this work, we additionally calculate $F$ for the simulations in the CAMELS-IllustrisTNG extended 1P28 set for which there are available spread metric data and the cosmic variance (CV) sets of the three suites. 

To confirm that the resolution of our 2D maps is sufficient to capture the DM variance in CAMELS, we perform a resolution test. We find that with $4000\times4000$, as we use, the results are converged and that increasing the resolution does not significantly alter our measured values of the F-parameter. As a second test, we plot the $P(DM \vert z)$ distributions and their fits according to Equation~\ref{eq:p_dm}. We find that the mean DM values fall within 3\% of the \cite{Walker_2023} measurements in IllustrisTNG-300 and that the standard deviation is significantly lower. This is what we expect due to the small CAMELS box size, as we discuss in further detail in Section~\ref{sec:disc}.

\subsection{The Baryon Spread Metric}

One way to capture the displacement of gas due to feedback is through the spread metric, introduced by \cite{Borrow_2020} with the SIMBA simulation. The spread metric is defined as the distance at the final conditions (usually $z = 0$) between that particle and the dark matter particle that it was closest to at initialization. Particles are matched between the initial conditions and the final snapshot using their particle IDs, which are consistent between snapshots. For IllustrisTNG, it is slightly more complicated, as there are gas cells that do not spread like particles but transfer mass between each other, so we use tracer particles as described in \cite{Genel_2013}. Upon simulation initialization, each gas cell is assigned a tracer particle that then evolves throughout the simulation as gas particles would, without carrying any mass. These tracer particles are tracked as gas particles and are used to compute the spread. For a hydrodynamic cosmological simulation, the gas spread is computed following the steps:

\begin{itemize}
    \item In the initial conditions, for each gas particle $i$, identify the $n$ nearest dark matter particle neighbors by computing the distance to each dark matter particle $j$. We store the particle IDs of these $n$ nearest dark matter particle neighbors.
    \item In the final snapshot, match each baryonic particle with its progenitor in the initial conditions using the particle ID. Star particles are matched with their gas progenitors. 
    \item For each gas particle $i$, using the stored IDs from step 1, find the positions of the $n$ nearest dark matter particles in the final conditions.
    \item In the final snapshot, calculate the distance between each baryonic particle $i$ and its nearest $n$ dark matter particle neighbors, $\rm r_{i,n}$. 
    \item The spread metric for particle $i$, given by $\rm S_i$, is equal to the median of the distances from the baryonic particle $i$ to the $n$ dark matter particles neighbors, $\rm r_{i,n}$.
\end{itemize}

A similar procedure can be followed to find the spread of dark matter particles. Thus, for a simulation, we can calculate the gas spread metric and the dark-matter spread metric. To characterize the spread in a simulation overall, we can calculate the normalized spread metric, which is the median baryon spread metric divided by the median dark-matter spread metric. 

The spread for a subset of simulations in the CAMELS project has been calculated and made publicly available by \cite{Gebhardt_2024}. This subset includes simulations in the CAMELS-SIMBA, CAMELS-Astrid, and CAMELS-IllustrisTNG. In the case of CAMELS-SIMBA and CAMELS-Astrid, star particles are not considered. See \cite{Gebhardt_2024} for more details on how the spread metric was calculated for CAMELS simulations. The spread metric data made available includes the spread of all gas and dark-matter particles in the simulation. Each particle is assigned to a host halo if the particle is within $R_{200c}$ of that halo at $z = 0$. We compute the median baryon spread, median dark matter spread, and normalized spread for all particles, particles outside of halos, and particles in halos.

\section{Results} \label{sec:res}

\subsection{The F-Parameter and Baryon Spread}

\begin{figure}
    \centering
    \includegraphics[width=\linewidth]{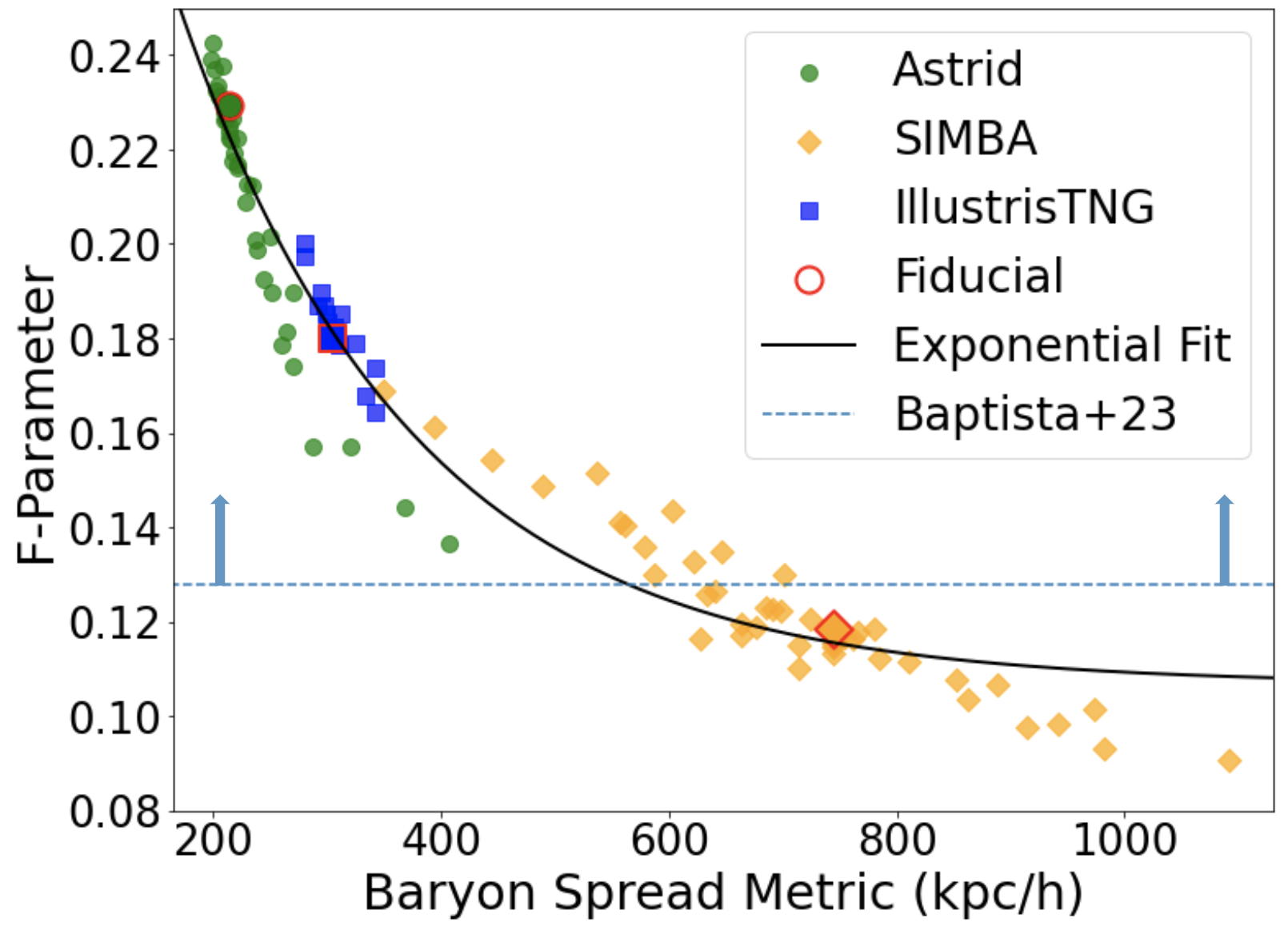}
    \caption{Median baryon spread (in units of kpc/h) versus the F-parameter (unitless) for simulations in CAMELS-Astrid (green circles), CAMELS-SIMBA (orange diamonds), and CAMELS-IllustrisTNG (blue squares), with the fiducial simulations outlined in red. The line of best fit in black is given by the exponential decay function: $F = 0.32841 \times e^{-0.00486 S} + 0.10679$. An observational lower limit measured by \cite{Baptista_2023} is marked by the blue-dashed line. \label{fig:spread-f}}
\end{figure}

By definition, $F$ measures the variance in DM from sightline to sightline at a given redshift, thus it should capture the degree globally baryons are pushed by feedback from halos. In Figure \ref{fig:spread-f} we plot the median baryon spread metric, $S$ over the entire box versus the F-parameter for simulations in the 1P sets CAMELS-Astrid (in green), CAMELS-SIMBA (in orange) and CAMELS-IllustrisTNG (in blue). Each point in the plot is a different simulation, with different subgrid physics. There are a total of 110 separate simulations: 44 CAMELS-SIMBA simulations, 44 CAMELS-Astrid simulations, and 22 CAMELS-IllustrisTNG simulations. These are from the 1P set of CAMELS where baryon spread data is available (limited to simulations with tracer particles). The fiducial version for each of the three suites is highlighted with a larger marker, outlined in red. 

\subsubsection{Feedback Effects}

Across all simulations, the values of the median baryon spread metric and the F-parameter span a wide range ($0.09 < F < 0.24$ and $198 < S < 1091$ kpc/$h$). The plot can be roughly divided into three regimes, each primarily containing the majority of simulations from one of the three suites: CAMELS-SIMBA, CAMELS-IllustrisTNG, and CAMELS-Astrid. Recall that $F$ scales inversely with the feedback strength, so that a value closer to zero indicates strong feedback and as $F$ approaches 1 it indicates weak feedback. In the lowest feedback regime, the values for $F$ are higher and the median baryon spread is lower. This region is primarily composed of simulations from the CAMELS-Astrid suite. The second regime is that of moderate feedback, where the CAMELS-IllustrisTNG simulations lie. Finally, the third regime is that of strong feedback (low $F$ values and a large median baryon spread), composed of CAMELS-SIMBA simulations. 

The division of the three suites into three regimes of feedback strength is in line with previous comparisons of subgrid models (both fiducial and within CAMELS). The effect of baryonic feedback can be characterized by looking at the ratio of the hydrodynamic simulation power spectrum compared to the power spectrum of its corresponding dark-matter-only simulations, $P_{\rm hydro}/P_{\rm nbody}$. If there were no effects on clustering due to baryonic physics, this ratio would be $P_{\rm hydro}/P_{\rm nbody} = 1$ on all scales. In practice, baryonic feedback suppresses $P_{\rm hydro}/P_{\rm nbody} < 1.0$ because it redistributes matter and suppresses clustering on large scales. At small scales ($k \gtrsim 10-20$ $h$/Mpc) the ratio of $P_{\rm hydro}/P_{\rm nbody} > 1$ due to increased clustering from the radiative cooling of baryons and star formation. In general, SIMBA has stronger effects on the matter power spectrum compared to IllustrisTNG \citep{camels_presentation, Gebhardt_2024, Delgado_2023, Panday_2023}. On the other hand, the Astrid subgrid model has the least impact on the matter power spectrum among the three suites \citep{camels_data_release2}. For example, the power spectrum ratio at z=0 can reach below $70\%$ on the $k \sim 10$ $h$/Mpc scale in SIMBA but only to $90\%$ in Astrid on the same scale. Another way of looking at the effect of feedback is by looking at how far baryons are pushed out of halos or baryon spread \citep{Gebhardt_2024}. SIMBA pushes baryons further out, with $\sim 40\%$ of baryons spreading to distances greater than 1 Mpc in the CAMELS-SIMBA fiducial model. In the fiducial CAMELS-Astrid model, only $7\%$ of baryons are pushed beyond 1 Mpc, and in CAMELS-IllustrisTNG a slightly higher percentage of $11\%$ is pushed beyond 1 Mpc. At intermediate scales, IllustrisTNG displaces significantly more baryons than Astrid. 

We can explain these differences by considering the differences in feedback implementation between the three suites. The difference in the large-scale gas distribution and matter power spectrum suppression is primarily caused by the difference in the AGN feedback models. Overall, AGN feedback is mildest in CAMELS-Astrid, strongest in CAMELS-SIMBA, and intermediate in CAMELS-IllustrisTNG. There are two key reasons we see this difference. First, in SIMBA, jet mode feedback is implemented as bipolar outflows. In contrast, in Astrid and IllustrisTNG kinetic and thermal energy are deposited locally, with no directionality. Thus, in SIMBA AGN feedback energy deposition reaches further distances, up to several times $R_{\rm 200c}$. In addition, these AGN-driven outflows can continuously entrain mass from the interstellar medium scale out to several hundreds of kpc \citep{Wright_2024}. The second key difference between the models is in the onset of feedback. Each of the three models has a different threshold mass to activate the kinetic mode: $M_{\rm BH} > 10^{7.5}$ in SIMBA, $M_{\rm BH} > 10^{8}$ in IllustrisTNG, and $M_{\rm BH} > 5 \times 10^{8}$ in Astrid. Thus, the onset of kinetic AGN feedback and thus the onset of matter power suppression due to baryonic effects occurs earliest in SIMBA and last in Astrid \citep{camels_data_release2}. In addition, we also must consider that the small box size of CAMELS inhibits the formation of the most massive structures that would display the strongest AGN jet feedback. As SIMBA has a lower mass threshold, it feels the effect of this suppression the least. For a detailed exploration of the effect of feedback implementation and parameter variations in the CAMELS-SIMBA and CAMELS-IllustrisTNG 1P sets on feedback energetics and CGM properties, see \cite{Medlock_2024b}. 

\subsubsection{F-Parameter and Baryon Spread Relation}

Notably, despite these significant differences in the subgrid physics implementation between the three models, and the effects on the baryon distribution, we observe that baryon spread and the F-parameter are tightly correlated, independent of the subgrid model. We are interested in baryon spread, as \cite{Gebhardt_2024} demonstrated a clear relationship between the impact of feedback on the matter power spectrum and the spread metric, in that greater spread correlates with a greater reduction of power. This holds even at small scales of $k \sim 30$ $h$/Mpc. The fractional difference in the total power spectrum due to baryonic effects can be expressed as a function of k and the spread, S, 
\begin{equation}
    -\Delta P/P = a_1 \times (k \times (a_2 \times S - a_a))^{1/4} -a_4,
\end{equation}
\noindent where for CAMELS-SIMBA \cite{Gebhardt_2024} found parameter fits of $a_1 = 0.25$, $a_2 =  0.35$, $a_3 = 1.09$, and $a_4 = 0.14$ with PySR symbolic regression \citep{miles_cranmer_2020_4041459}. While this is a promising avenue for correcting the matter power spectrum for baryonic effects, the spread metric is something that is only quantifiable via simulations and cannot be measured observationally. The tight correlation between the observable quantity $F$ and the spread metric that we show in Figure \ref{fig:spread-f}, demonstrates that the F-parameter can be used as an observational proxy for baryon spread to assess the degree of suppression in the matter power spectrum due to baryonic feedback physics.

\subsubsection{Analytical Fit}

We fit the correlation between median baryon spread and the F-parameter with an exponential decay function. The exponential decay fit was the best of all the functions we tried, including polynomials. The best fit is shown in black in Figure \ref{fig:spread-f} and given by
\begin{equation} \label{f_s}
    F = 0.32841 \times e^{-0.00486 S} + 0.10679,
\end{equation}
\noindent where $S$ is the median baryon spread metric for all baryonic particles in a simulation with units of kpc/$h$ and $F$ is the F-parameter for the same simulation. Using this relation, we can estimate the median baryon spread of our Universe with a measurement of the F-parameter. While this choice was initially arbitrary, the relatively good fit of the exponential decay function suggests there may be some physical argument behind the relation, whose understanding is beyond the scope of this work but is worthy of future investigation.

\subsubsection{Comparison to Observations}

In the last few years, a couple of preliminary observations of the F-parameter have been made. First,  \cite{Macquart_2020} measured $F = 0.23^{+0.27}_{-0.12}$ with 7 FRBs. More recently, \cite{Baptista_2023} found measured $F = 0.33^{+0.27}_{-0.11}$, and a lower limit of $F > 0.128$ with 99.7\% confidence, with 78 FRBs of which 21 were localized. According to our exponential decay fit, this roughly corresponds to a maximum median baryon spread of $\sim 563.74$ kpc/h. Just considering the simulation with an $F$ above our lower limit, the maximum median baryon spread is $\sim 700.77$ kpc/h. However, these current measurements of the F-parameter have large errors due to the small sample of localized FRBs (FRBs with precise enough spatial and redshift localization to map to a host halo and assign a redshift). \cite{Baptista_2023} forecasts with synthetic FRBs that 100 localized FRBs will be sufficient to constrain the upper and lower limits on $F$ to $3 \sigma$. Although we do not yet have such a sample, in the next couple of years, thanks to current and planned facilities like CHIME outriggers, CHORD, DSA-2000, and BURSTT we expect to detect around $10,000$ of FRBs with host galaxies per year \citep{Leung_2021, Vanderline_2019, Hallinan_2019, BURSTT_2022}. In addition, we would like to note that we don't expect our F-parameter values from CAMELS to be comparable to real measurements yet. This comparison would require further care and attention to systematics and assumptions. Independent of this comparison, this theoretical result is notable and demonstrates the need for further work so it can be applied to observations in the future.

\begin{figure}
    \centering
    \includegraphics[width=\linewidth]{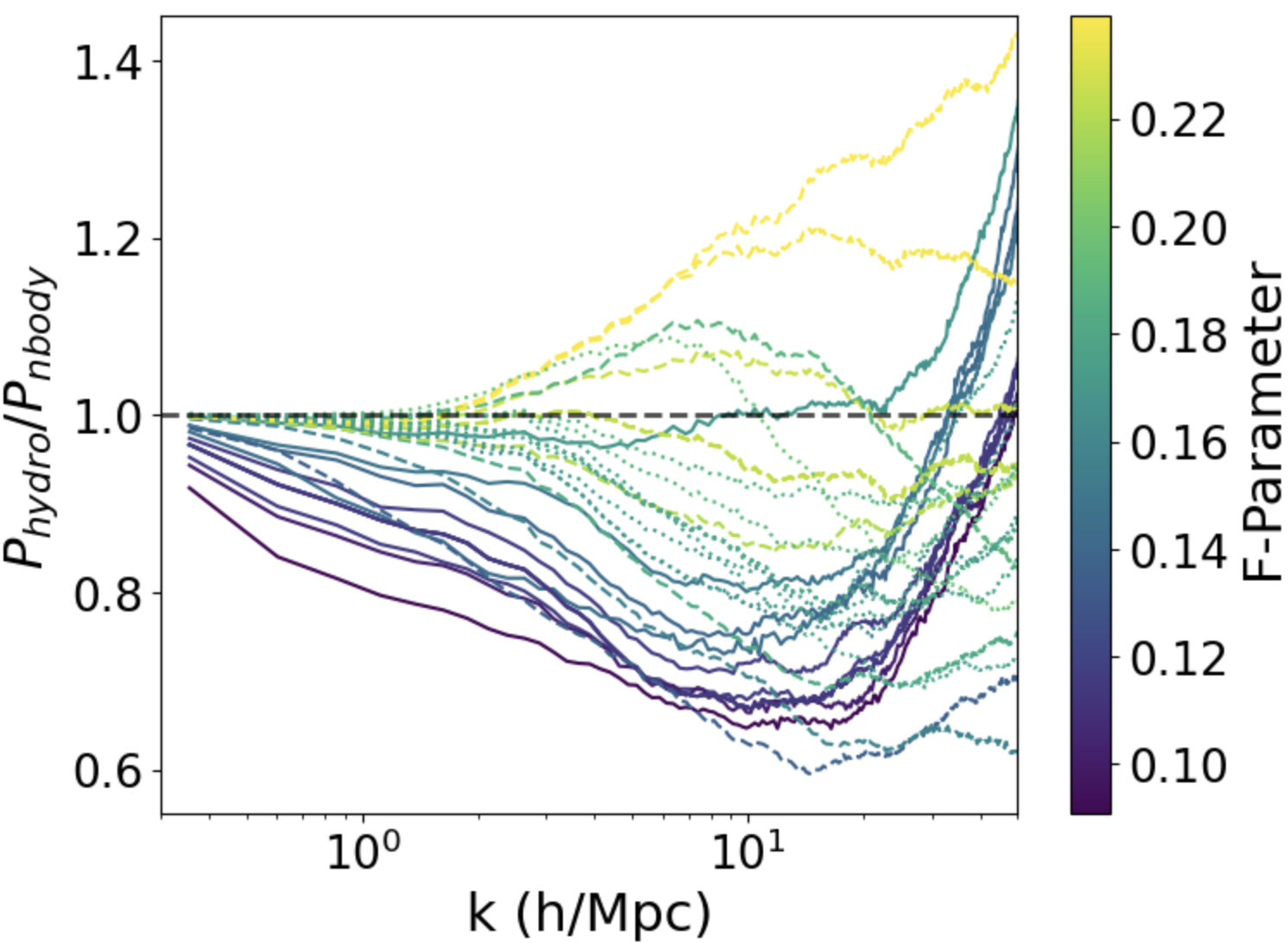}
    \caption{The ratio of hydrodynamic versus N-body total matter power spectra for simulations in the 1P sets of CAMELS-SIMBA (solid lines), CAMELS-IllustrisTNG (dotted lines), and CAMELS-Astrid (dashed lines), colored by corresponding F-parameter value. \label{fig:pow_spec}}
\end{figure}

\subsection{F-Parameter and Matter Power Suppression}

Alternatively, we explore using the F-parameter directly to constrain the effect of feedback on the matter power spectrum. Emulating \cite{Gebhardt_2024}, in Figure \ref{fig:pow_spec} we plot the total ratio of the total matter power spectrum of the hydrodynamic simulation run divided by that of the corresponding dark matter-only simulation ($P_{\rm hydro}/P_{\rm nbody}$). $P_{\rm nbody}$ is the power spectrum from the N-body (dark matter-only) simulation corresponding to each hydrodynamical simulation. If there were no baryonic effects, then this ratio would be zero, denoted with a black dashed line. We plotted this ratio for 36 total distinct simulations in the CAMELS-SIMBA (solid lines), CAMELS-IllustrisTNG (dotted lines) and CAMELS-Astrid (dashed lines) 1P sets. Each line is colored by the corresponding F-parameter value. We clearly see a trend in which simulations that display greater suppression due to feedback correspond to lower F-parameter values. The trend appears to remain straightforward on scales of $k < 10$ $h$/Mpc. After this point the simulations become entagled, due to complex nonlinear effects.

\begin{figure}
    \centering
    \includegraphics[width=\linewidth]{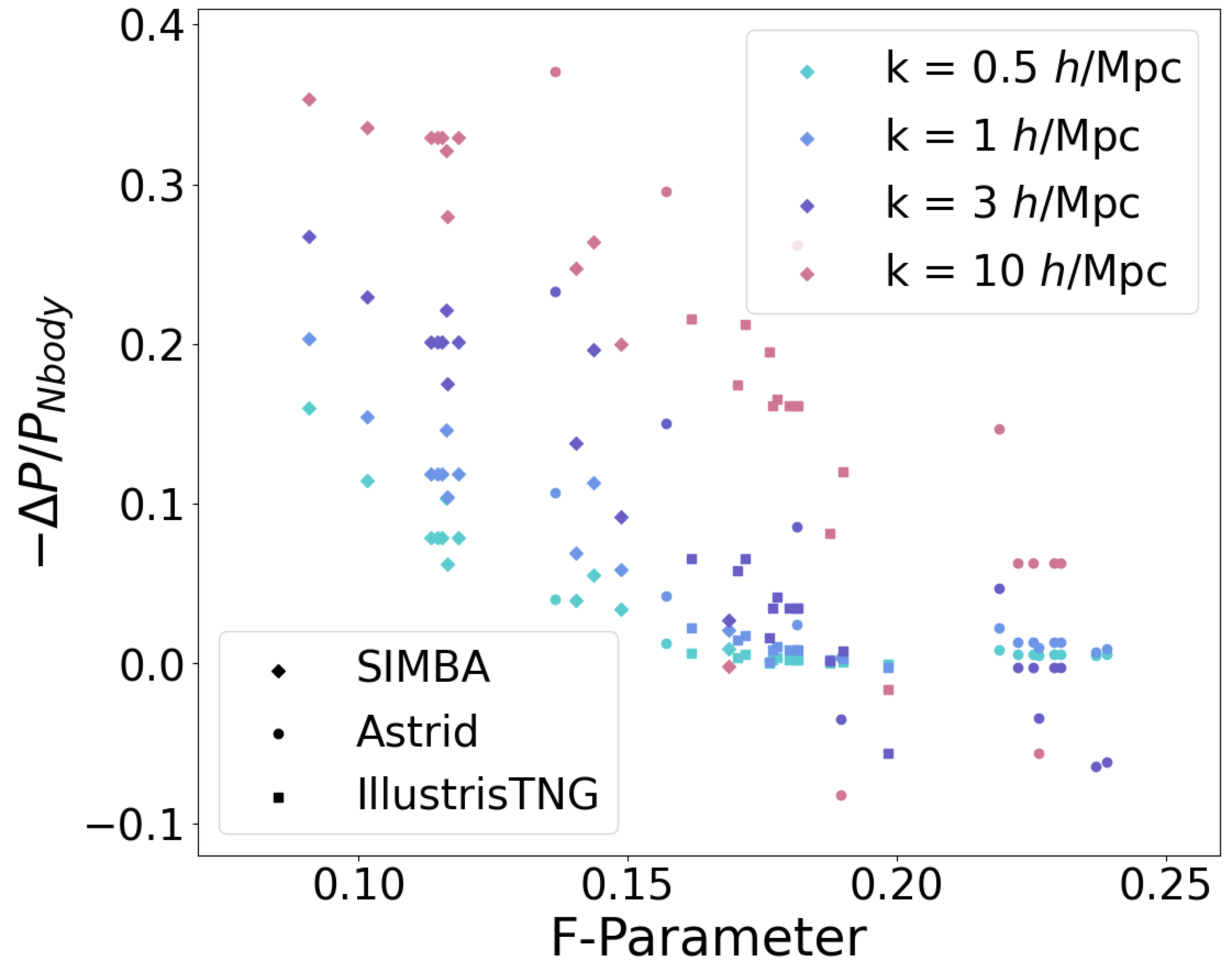}
    \caption{The fractional power difference at $z=0$ as a function of the F-parameter at different values of $k$. Simulations from CAMELS-SIMBA are plotted as diamonds, CAMELS-IllustrisTNG as squared, and CAMELS-Astrid as circles. \label{fig:frac_diff}}
\end{figure}

To further elucidate the trend for a given scale, in Figure \ref{fig:frac_diff} we plot the fractional power difference for four different scales ($k = 0.5$, $1$, $3$, and $10$ $h/Mpc$). We define fractional power difference by

\begin{equation}
    -\frac{\Delta P}{P_{\rm nbody}} = - \frac{P_{\rm hydro} - P_{\rm nbody}}{P_{\rm nbody}},
\end{equation}

\noindent where $P_{\rm hydro}$ is the total matter power spectrum from the hydrodynamic run of a simulation and $P_{\rm nbody}$ is the total matter power spectrum from the corresponding dark matter only N-body simulation. As expected, we observe a negative trend between the F-parameter and the fractional power difference, with an offset depending on the scale $k$.

\section{Discussion} \label{sec:disc}

\subsection{Caveats and Limitations}

Although there is a remarkable correlation between the F-parameter and the spread metric, some scatter is present. To understand the origin and mitigate the effects of this scatter, we explore the dependence of the relationship on several factors. First, we investigate the difference in behavior between particles that have a host halo and said host halo's mass compared to those that are outside of the halos. Second, we examine the subgrid physics of individual models that deviate the most from the relation. Third, we consider the accuracy of our assumption that the F-parameter is constant for all redshift. Lastly, we consider the effect of cosmic variance and the size of the CAMELS project simulation boxes.

\subsubsection{Baryon Spread and Host Halos}

A particle is considered inside a halo if at $z = 0$ the particle is within $R_{\rm 200c}$ of the halo. All particles that do not meet this criteria are designated as particles outside the halos. As shown in \cite{Gebhardt_2024}, dark matter particles spread further when outside of a halo than when associated with a halo. We find that baryons outside of halos primarily drive the relation we see in Figure \ref{fig:spread-f}, shifted slightly toward higher spread. Baryons with host halos follow roughly the same relationship, shifted significantly lower in spread, except for CAMELS-SIMBA and a few simulations in CAMELS-Astrid, for which we observe the opposite correlation ($S$ correlates positively with $F$). This is likely due to how particles are assigned a host halo. The designation of particles in a host halo is only capturing particles that have not spread very far, not those that started in the halo but were ejected beyond because of strong feedback. In addition, focusing on baryons with host halos, we bin the particles by host halo mass. We find that for baryons with host halos of $M_{\rm h} < 10^{12.5} M_{\odot}$ the spread remains practically uniform within all 110 simulations in the three subgrid models. Particles with host halos with $M_{\rm h} > 10^{12.5} M_{\odot}$ show the same negative correlation between the spread metric and the F-parameter, both when considering the baryon spread metric and the normalized spread metric. 

\subsubsection{Deviations from the S-F Relation}

Next, we examine individual simulations that deviate the most from the correlation. The three simulations that deviate the most are from CAMELS-Astrid. These correspond to the simulations with the lowest three values of the parameter $A_{\rm AGN1}$. In \cite{Medlock_2024}, we observed that variations of $A_{\rm AGN1}$ in CAMELS-Astrid exhibited unique behavior compared to any parameter in any suite. The effect of this parameter, which controls jet-mode feedback and turns on at the highest threshold of the three suites ($5 \times 10^{8} M_{\odot}$) is not fully captured in the simulations because the small box size does not allow the formation of these most massive objects that trigger jet mode feedback \citep{camels_data_release2}. To fully understand the effect of this mode of feedback, we again need to repeat this analysis on larger simulations. While we focus on the effects of baryonic feedback prescriptions on the S-F relation for the CAMELS-IllustrisTNG suite, we analyzed some simulations that vary cosmology (one of three parameters: $\rm \Omega_b$, $H_0$, and $n_{\rm e}$). These points are not included in Figure~\ref{fig:spread-f}, where the points all have fixed cosmology. \cite{Medlock_2024} showed that the F-parameter is sensitive to large-scale structure and cosmology, specifically to $\rm \Omega_m$ and $\sigma_8$, where an increase in either parameter leads to an increase in the F-parameter. When we include simulations with varied cosmology in the S-F relation, we found that the normalization is shifted up or down the y-axis (the F-parameter). This is what we expect, for the quantities to remain correlated, but with a different fit due to the shift in values depending on cosmology. A full understanding of the cosmological effects would require further analysis of simulations with both varied cosmologies and baryonic feedback, for example using the LH set of CAMELS.

\subsubsection{F-parameter Redshift Dependence}

\begin{figure}
    \centering
    \includegraphics[width=\linewidth]{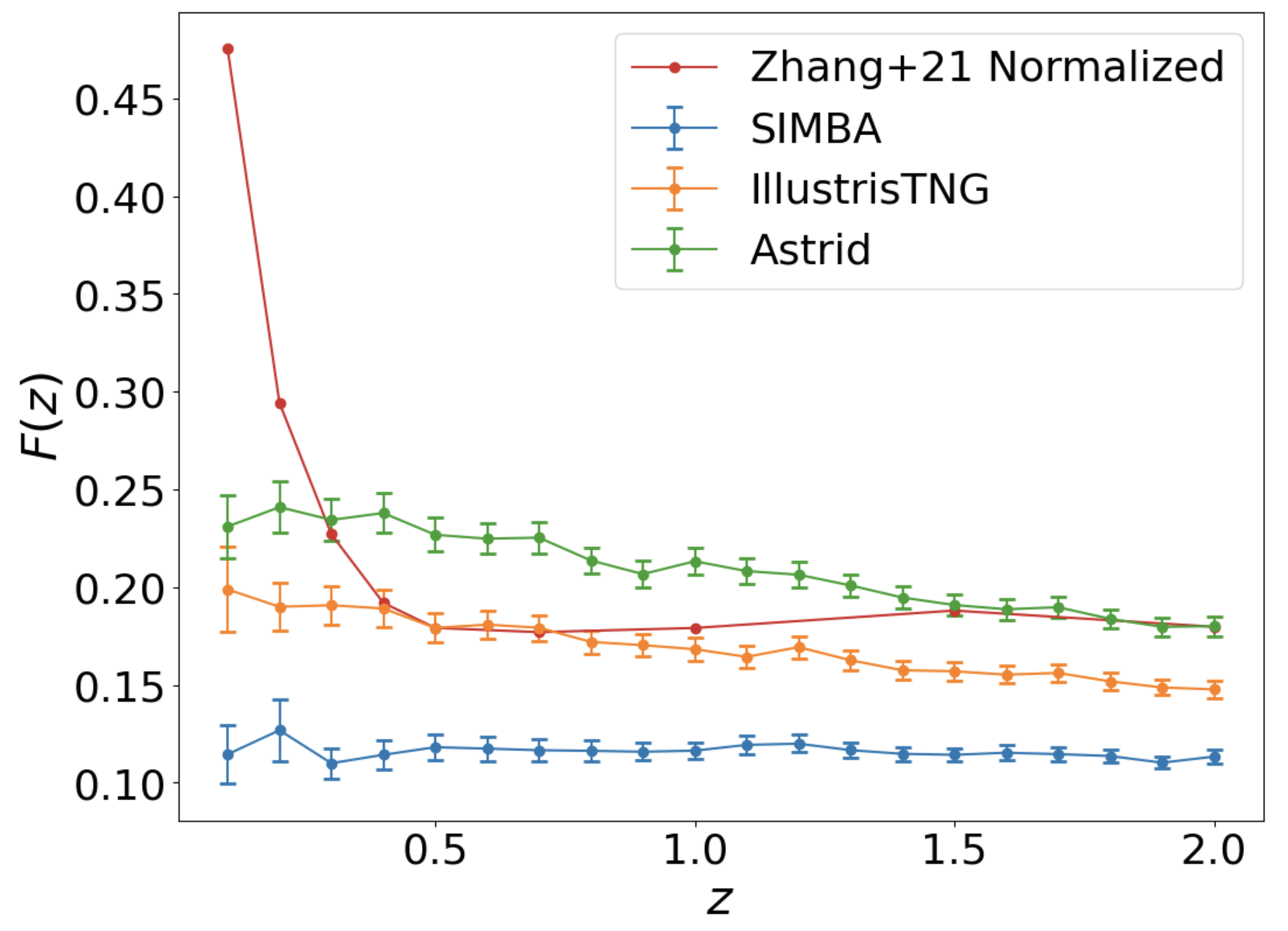}
    \caption{The redshift evolution from $z = 0.1 - 2$  of the F-parameter for the CAMELS-SIMBA (in blue), CAMELS-IllustrisTNG (in orange), and CAMELS-Astrid (in green) fiducial models. We compare against the \cite{Zhang_2021} $F$-z relation (in red) calculated with IllustrisTNG-300, normalized so that the values of both IllustrisTNG models match at $z = 0.5$. Error bars for the CAMELS suites indicate the 1$\sigma$ error range.}
    \label{fig:Fz}
\end{figure}

One key assumption that underlies this relation is that the F-parameter is constant as a function of the redshift (specifically, for our relation, we assume that $F$ is constant between $z=0-0.5$). This assumption is consistent with those of previous work \citep{James_2022, Baptista_2023}. However, it is not necessarily true. For example, \cite{Zhang_2021} examined the redshift dependence of $F$ in IllustrisTNG and found that $F$ is constant for $0.4 < z < 2.0$ but in most other regimes it decreases as a function of $z$. In general, there has been little work done to characterize the redshift dependence of the F-parameter. To characterize the evolution of the F parameter in CAMELS, we calculate $F$ at various redshifts of $z = 0 - 2$ for the fiducial models of CAMELS-SIMBA, CAMELS-IllustrisTNG, and CAMELS-Astrid. These results are presented in Figure \ref{fig:Fz} along with the results from \cite{Zhang_2021}, normalized so that the F-parameter values of IllustrisTNG-300 and CAMELS-IllustrisTNG match at $z = 0.5$.

For CAMELS-IllustrisTNG and CAMELS-Astrid we find a slight decrease in the F parameter as a function of the redshift, while in CAMELS-SIMBA $F$ remains relatively constant throughout the $z$ range. We note a significant discrepancy between the behavior of the F-parameter in CAMELS-IllustrisTNG and IllustrisTNG-300, especially at $z < 0.5$. In IllustrisTNG-300 we observe a sharp increase in $F$ as $z$ approaches zero, while in CAMELS-IllustrisTNG $F$ remains relatively constant. It is not clear whether the CAMELS boxes can accurately capture the true evolution of the F-parameter due to the limited volume, necessitating further study.

Furthermore, we examine the evolution of $F$ across $z=0.25-1$ for the entire 1P sets of CAMELS-SIMBA, CAMELS-IllustrisTNG, and CAMELS-Astrid. We find that in this redshift range $F$ is relatively constant with a mean deviation from $F(z=0.5)$ of 2.3\% for $F(z=0.25)$, 2.2\% for $F(z=0.75)$ and 3.0\% for $F(z=1.0)$. The maximum deviation is within 10\%, which is within $1\sigma$ error of our $F$ measurements.

\subsubsection{Cosmic Variance and Simulation Box Size}

\begin{figure*}
    \centering
     \includegraphics[width=\linewidth]{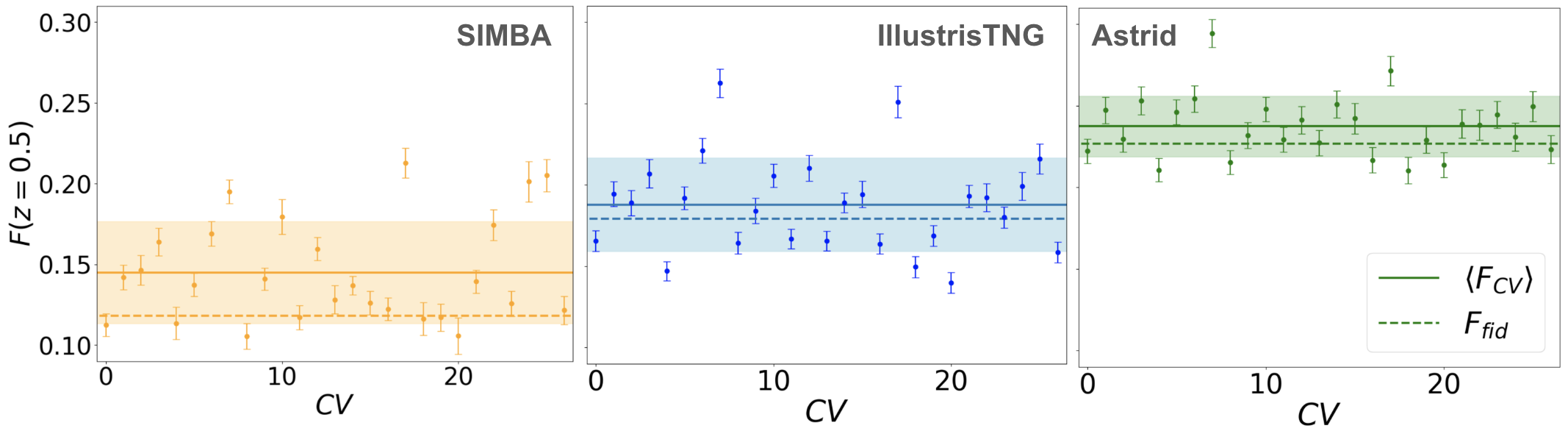}
    \caption{F-parameter values of the simulations of the cosmic variance sets from CAMELS-SIMBA (left panel), CAMELS-IllustrisTNG (middle panel), and CAMELS-Astrid (right panel). The mean value is marked with the solid blue line with the 1$\sigma$ region shaded in blue. The fiducial value from the 1P set is shown with the dashed line.}
    \label{fig:CV}
\end{figure*}

\begin{figure*}
    \centering
     \includegraphics[width=\linewidth]{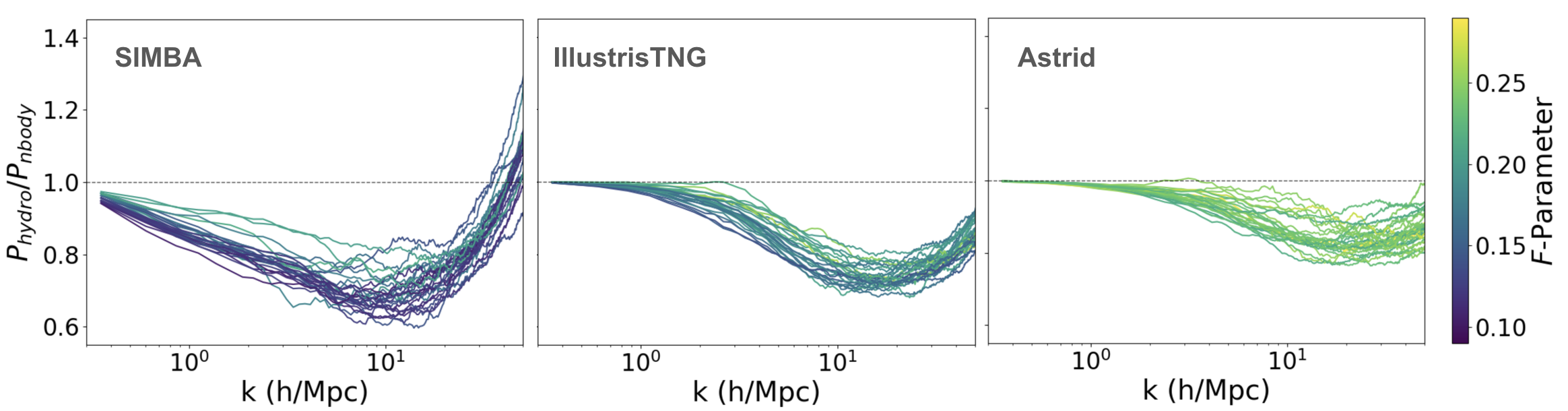}
    \caption{The ratio of hydrodynamic versus N-body total matter power spectrum for simulations in the CV sets of  CAMELS-SIMBA (left panel), CAMELS-IllustrisTNG (middle panel), and CAMELS-Astrid (right panel), colored by corresponding F-parameter value. }
    \label{fig:CV_Pk}
\end{figure*}

Lastly, when applying these results, we need to consider the effects the small box size of the CAMELS project simulations has on the results. In CAMELS it has been shown that cosmic variance alone represents a significant amount of scatter in power suppression even for fiducial feedback parameters \citep{camels_presentation, Delgado_2023, camels_data_release2}. \cite{Gebhardt_2024} notes that there is significant scatter in the relation between fractional power suppression and normalized spread due to cosmic variance. To assess this effect on the F-parameter we use the CAMELS cosmic variance sets to estimate how much scatter in measurements of the F-parameter we get using different random initial seeds. The F-parameter measurements at $z=0.5$ for CAMELS-SIMBA, CAMELS-IllustrisTNG, and CAMELS-Astrid are shown in Figure \ref{fig:CV}. Here we see that for CAMELS-IllustrisTNG, the variation due to the change of initial seeds is significantly greater than the dependence on $z$. In Figure \ref{fig:CV_Pk} we plot $P_{\rm hydro}/P_{\rm nbody}$ for the corresponding CV set simulations. These plots demonstrate that correlation between the degree of suppression and the F-parameter is still clear despite cosmic variance.

In addition, there is a systematic error in the measurements of $F$ in CAMELS. Due to the small box size, the measurements of $F$ are biased to lower values because the simulations do not capture the effect of the large-scale structure, voids, and clusters that add significantly more variation to the DM-z relation. Consequently, the matter power spectrum is not properly normalized on all scales \citep{camels_presentation}. Furthermore, \cite{vanDaalen_2020} showed that halos with $M_{h} > 10^{14}$ dominate the power on scales of $k > 1$ $h$/Mpc. However, since CAMELS is a small box, there are only a few halos with $10^{12} M_{\odot} < M_h < 10^{14} M_{\odot}$. This is another reason to repeat the analysis on larger boxes that contain a sufficient number of high-mass halos that, in practice, have the greatest effect. Our current results show that there is a strong relation comparatively within suites but to robustly characterize this relation and apply it to observations it is necessary to continue this study with larger simulation boxes that are converged with respect to dispersion measure variance and cosmic variance. Again, we would like to note that we don't expect our F-parameter values from CAMELS to be comparable to real measurements yet. This is primarily a proof of concept, so as long as they are comparable to each other, the main conclusion of the F-parameter being a potential tool to constrain baryonic effects on the matter power spectrum still holds.

\subsection{Future Directions}

Based on the potential demonstrated by these results, we propose to develop a framework in the coming years to use FRBs to calibrate the matter power spectrum for more precise measurements of $S_{8}$ with weak lensing. In order to do this, we need to calibrate the relation of matter power suppression and the F-parameter with sufficiently large enough simulations. Other quantities we can consider in addition to baryon spread are the retained halo baryon fraction, $f_{\rm b}$, which \cite{vanLoon_2024} showed is a good predictor of power suppression, and the closure radius, which \cite{Ayromlou_2023} shows is a good tracer of baryon redistribution. Both of these quantities are challenging or impossible to measure observationally and necessitate a simple proxy like the F-parameter. In addition, we need to consider statistics beyond the F-parameter, which despite the potential shown here, is a simplification of the available data. One possibility is the baryon fraction in the IGM ($f_{\rm IGM}$) which is sensitive to the effects of feedback and has recently been measured by a sample of localized FRBs in \cite{Connor_2024}.

Ultimately, the goal is to incorporate FRB constraints into a multiprobe, multiwavelength approach to correct the matter power spectrum and better constrain $S_{8}$. Recently, \cite{Reischke_2023} proposed the cross-correlating of the FRB dispersion measure with the weak lensing signal to correct for baryonic feedback, showing that $\sim 10^4$ FRBs correlated with a weak lensing signal from a Euclid-like survey would constrain the feedback 10 times better than through cosmic shear alone. \cite{Nicola_2022} uses CAMELS to show that the electron auto-power spectrum, which is measurable with FRBs and through the kinematic SZ observations, provides tight constraints on cosmological and astrophysical parameters, largely robustly to the differences in the subgrid model. These examples highlight the potential of FRBs as both an independent and complementary probe to yield more accurate constraints on cosmological and astrophysical parameters, ultimately leading to a deeper understanding of baryonic feedback mechanisms and the underlying structure of the Universe. In the next years, in preparation for upcoming localized FRB observations, we need to address concerns of the redshift dependence of the F-parameter, the uncertainties and in the dispersion measure contributions of the Milky Way and the host halos of FRBs, and prepare predictions of the F-parameter for various models that emulate observations as closely as possible.

To address some of these concerns, we plan to repeat this analysis on the ongoing and upcoming 50~Mpc/h CAMELS boxes, as well as the original versions of Astrid, SIMBA, and IllustrisTNG. In addition we can add new simulations such as FLAMINGO \citep{Flamingo_2023} which presents larger boxes with varied feedback implementations, the different feedback implementations of SIMBA, and additional models like RAMSES \citep{ramses_2002}, ENZO \citep{enzo_2014}, and Magneticum \citep{magneticum_2016} to test robustness to subgrid model. We can also repeat this analysis on the Latin-Hypercube CAMELS sets to look at the impact of cosmology and feedback effects simultaneously. In order to prepare to compare directly to observations, it is also necessary to increase the robustness of FRB predictions. We need to better understand the redshift dependence of the F parameter in both simulations and observations. It is also worth considering how we construct mock FRB sightlines, including where they begin in a simulation box, and how we assign an electron density value along each point in the sightline. As noted in \cite{Guo_2025}, differences in computational techniques can lead to significantly different measurements of DM on the order of 10\%-20\%, as well as the behavior of the F-parameter, even within the same simulations. 

\section{Conclusions} \label{sec:conc}

With the advent of powerful, large-scale galaxy and lensing surveys such as Rubin \citep{Ivezic_2019A}, Roman \citep{WFIRST_2015}, and Euclid \citep{Euclid_2022} we are entering an era of precision cosmology. However, to maximize the statistical power of these surveys we must control the uncertainties related to the poorly understood baryonic feedback effects on the matter power spectrum. Specifically, to address systematic uncertainties in weak lensing measurements of $S_{8}$ we need to accurately correct for baryonic effects on the matter power spectrum. 

We propose a novel method for this correction with fast radio burst (FRB) observables. In particular, we focus on the F-parameter, which describes the degree of variation in the dispersion measures of FRB sightlines at a given redshift. Our previous work has shown that the F-parameter captures the difference in the baryon distribution between different subgrid models in the CAMELS project. In this work, we explore the potential use of the F-parameter to constrain cosmology, specifically $S_{8}$. We summarize our findings as follows.

\begin{itemize}
    \item The F-parameter strongly correlates, independent of the subgrid model, with the baryon spread metric, a quantity calculable in simulations that has been shown to trace the effects of baryonic feedback on the matter power spectrum (Fig. \ref{fig:spread-f}). 

    \item The relation between the F-parameter and the baryon spread metric (the $S$-$F$ relation) can be fit analytically with an exponential decay function (Fig. \ref{fig:spread-f}). We propose to use this relation on future large populations of FRBs to constrain the impact of baryonic physics on the matter power spectrum. 

    \item The F-parameter can be used directly to capture the effects of baryons on the matter power spectrum, as it correlates strongly with overall power suppression (Fig. \ref{fig:pow_spec}). We can characterize the degree of fractional power suppression as a function of scale, $k$, and the F-parameter (Fig. \ref{fig:frac_diff}).

    \item Several uncertainties remain to be addressed before these relations can be applied to observations. In particular, we need to understand the redshift evolution of FRB observables such as the F-parameter (Fig. \ref{fig:Fz}) as well as the effect of cosmic variance in simulation predictions (Fig. \ref{fig:CV} and Fig. \ref{fig:CV_Pk}).
\end{itemize}

We have demonstrated that FRBs are a potentially powerful and clean way to characterize the effect of baryons on the matter power spectrum. However, there are still many uncertainties and caveats that need to be understood in order to apply to upcoming FRB observations. In the next couple of years we expect to have sufficient FRBs with host galaxies to constrain cosmology. Thus, now is the time to harness advances in simulations to prepare these theoretical predictions.

\section*{Acknowledgements}
The authors gratefully acknowledge the CAMELS simulations, which were performed on the supercomputing facilities of the Flatiron Institute supported by the Simons Foundation. This work is supported by the NSF grant AST 2206055 and the Yale Center for Research Computing facilities and staff. IM acknowledges support from the Dean's Emerging Scholars Research Award from the Yale Graduate School of Arts \& Sciences. DAA acknowledges support by NSF grant AST-2108944, NASA grant ATP23-0156, STScI grants JWST-GO-01712.009-A, JWST-AR-04357.001-A, and JWST-AR-05366.005-A, and Cottrell Scholar Award CS-CSA-2023-028 by the Research Corporation for Science Advancement.

\bibliography{main}
\bibliographystyle{aasjournal}

\appendix
\section{Dispersion Measure Probability Distributions for the Fiducial Models}

\begin{figure*}[h]
    \centering
     \includegraphics[width=\linewidth]{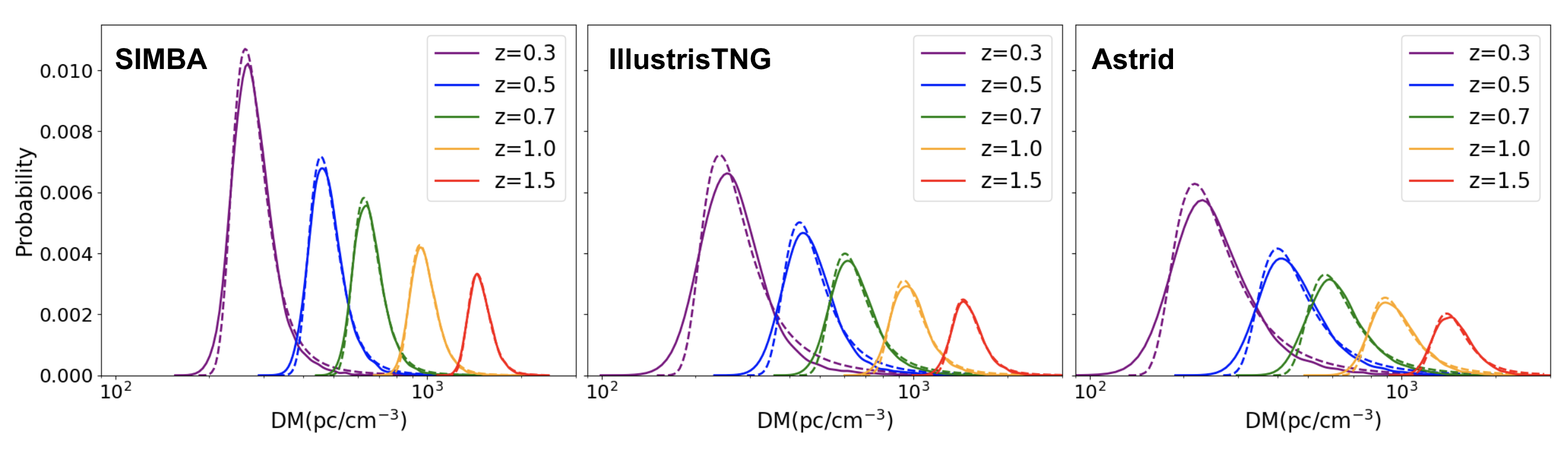}
    \caption{We show the $\rm P(DM\mid z)$ distributions from the fiducial models of CAMELS-SIMBA (left), CAMELS-IllustrisTNG (middle), and CAMELS-Astrid (right) for $z = 0.3$ (purple), $z = 0.5$ (blue), $z = 0.7$ (green), $z = 1.0$ (yellow), and $z = 1.5$ (red). \textbf{The dashed are the lines of best fit to Equation~\ref{eq:p_dm} scaled from $\rm P(\Delta \mid z)$ to $\rm P(DM\mid z)$.}}
    \label{fig:P_DM}
\end{figure*}

For robustness and reproducibility, we plot the $\rm P(DM\mid z)$ distributions for $z \in [0.3, 0.5, 0.7, 1.0, 1.5]$ in Figure~\ref{fig:P_DM}. These distributions are depicted with solid lines. \textbf{We fit these distributions according to Equation~\ref{eq:p_dm}, which fits $\rm P(\Delta \mid z)$, where $\Delta = \rm DM/\langle DM \rangle$. In Figure~\ref{fig:P_DM}, the dashed lines correspond to these fits transformed to $\rm P(DM)$. Table~\ref{tab:P_DM} provides the $\rm \langle DM \rangle$ values calculated directly from the data distributions, along with the parameters for the lines of best fit of $\rm P(\Delta \mid z)$.}

\begin{table}[ht]
\centering
\begin{tabular}{c|ccc|c|ccc|c|ccc|c}
\toprule
 & \multicolumn{4}{c|}{\textbf{CAMELS-SIMBA}} & \multicolumn{4}{c|}{\textbf{CAMELS-IllustrisTNG}} & \multicolumn{4}{c}{\textbf{CAMELS-Astrid}} \\
\midrule
\textbf{z} & \( A \) & \( \sigma \) & \( C_0 \) & \( \langle \text{DM} \rangle \) (pc/cm\(^3\)) & \( A \) & \( \sigma \) & \( C_0 \) & \( \langle \text{DM} \rangle \) & \( A \) & \( \sigma \) & \( C_0 \) & \( \langle \text{DM} \rangle \) \\
\midrule
\textbf{0.3} & \( 2.430 \) & \( 0.172 \) & \( 1.181 \) & \( 290.59 \) & \( 1.326 \) & \( 0.377 \) & \( 1.265 \) & \( 297.82 \) & \( 0.974 \) & \( 0.663 \) & \( 1.055 \) & \( 298.28 \) \\
\textbf{0.5} & \( 2.930 \) & \( 0.140 \) & \( 1.144 \) & \( 496.11 \) & \( 1.778 \) & \( 0.250 \) & \( 1.232 \) & \( 503.06 \) & \( 1.263 \) & \( 0.407 \) & \( 1.256 \) & \( 504.67 \) \\
\textbf{0.7} & \( 3.341 \) & \( 0.122 \) & \( 1.113 \) & \( 671.17 \) & \( 2.037 \) & \( 0.212 \) & \( 1.204 \) & \( 687.25 \) & \( 1.497 \) & \( 0.316 \) & \( 1.241 \) & \( 686.40 \) \\
\textbf{1.0} & \( 3.786 \) & \( 0.107 \) & \( 1.085 \) & \( 1001.27 \) & \( 2.573 \) & \( 0.162 \) & \( 1.144 \) & \( 1022.47 \) & \( 1.891 \) & \( 0.234 \) & \( 1.204 \) & \( 1020.49 \) \\
\textbf{1.5} & \( 4.575 \) & \( 0.088 \) & \( 1.062 \) & \( 1503.51 \) & \( 3.313 \) & \( 0.123 \) & \( 1.094 \) & \( 1536.31 \) & \( 2.542 \) & \( 0.167 \) & \( 1.116 \) & \( 1525.57 \) \\
\bottomrule
\end{tabular}
\caption{We present the mean DM (calculated directly from the distribution) as well as the best fit parameters of the $\rm P(\Delta \mid z)$ distributions for $z \in [0.3, 0.5, 0.7, 1.0, 1.5]$ for the fiducial models of CAMELS-SIMBA, CAMELS-IllustrisTNG, and CAMELS-Astrid, corresponding to those plotted in Figure \ref{fig:P_DM}.}
\label{tab:P_DM}
\end{table}

\end{document}